\documentclass[superscriptaddress,twocolumn,showpacs,prb]{revtex4}

\usepackage{graphicx,amsmath}
\usepackage{dcolumn}
\usepackage{amssymb}
\usepackage{bm}
\usepackage{color}
\newcommand{\av}{_\text{av}}
\newcommand{\bfk}{\boldsymbol{k}}

\begin{document}

\title{Phase transition in the three dimensional Heisenberg spin glass:\\
   Finite-size scaling analysis
}

\author{L.~A.~Fernandez}
\affiliation{Departamento
de F\'\i{}sica Te\'orica I, Universidad
Complutense, 28040 Madrid, Spain.}
\affiliation{Instituto de Biocomputaci\'on y
F\'{\i}sica de Sistemas Complejos (BIFI), Zaragoza, Spain.}

\author{V.~Martin-Mayor}
\affiliation{Departamento de F\'\i{}sica
Te\'orica I, Universidad Complutense, 28040 Madrid, Spain.}
\affiliation{Instituto de Biocomputaci\'on y 
F\'{\i}sica de Sistemas Complejos (BIFI), Zaragoza, Spain.}

\author{S.~Perez-Gaviro}
\affiliation{Dipartimento di Fisica, SMC of INFM-CNR and
  INFN, Universit\`a di Roma {\it La Sapienza}, 00185 Roma, Italy.}
\affiliation{Instituto de Biocomputaci\'on y
F\'{\i}sica de Sistemas Complejos (BIFI), Zaragoza, Spain.}

\author{A.~Tarancon}
\affiliation{Instituto de Biocomputaci\'on y
F\'{\i}sica de Sistemas Complejos (BIFI), Zaragoza, Spain.}
\affiliation{Departamento
de F\'\i{}sica Te\'orica, Universidad
de Zaragoza, 50009 Zaragoza, Spain.} 

\author{A.~P.~Young}
%\email{peter@physics.ucsc.edu}
\affiliation{Department of Physics, University of California,
Santa Cruz, California 95064}

\date{\today}

\begin{abstract}
We have investigated the phase transition in the Heisenberg spin glass
using massive numerical simulations to study larger sizes, $48^3$, than
have been attempted before at a spin glass phase transition. A
finite-size scaling analysis indicates that the data is compatible with
the most economical scenario: a common transition temperature for spins
and chiralities.
\end{abstract}
\pacs{75.50.Lk, 75.40.Mg, 05.50.+q}
\maketitle

\section{Introduction}
As a result of extensive and careful numerical
studies,\cite{ballesteros:00,katzgraber:06,hasenbusch:08b}
there is now compelling evidence for a finite temperature phase
transition in the Ising spin glass in three dimensions. However, the
situation for the Heisenberg spin glass, in which the spins are
classical 3-component vectors, is still controversial. The Heisenberg
spin glass is a suitable first model to describe experimental systems
with weak anisotropy, such as dilute Mn atoms in Cu
which is a well studied spin glass system, see
e.g.~ Ref.~\onlinecite{omari:83}.
Kawamura~\cite{kawamura:98,hukushima:05}
proposed that the spin glass transition only occurs at $T_\text{SG} =
0$ and that a \textit{chiral} glass transition occurs at a finite
temperature $T_\text{CG}$. Chiralities are Ising-like variables which
describe the handedness of the non-collinear spin structure. This
scenario requires that spins and chiralities decouple at long length
scales. However
simulations~\cite{lee:03,campos:06,lee:07}
subsequently found evidence for a finite $T_\text{SG}$, though
corrections to the leading finite-size scaling behavior seem larger
than in the Ising case.\cite{hasenbusch:08b} Recently, Viet and
Kawamura~\cite{viet:09,viet:09a} who did a similar analysis to that of
Refs.~\onlinecite{campos:06,lee:07} and used the same range of sizes ($L\le
32$, where $L$ is the linear size of, the system), concluded that
$T_\text{SG}$ is indeed finite, but is less than $T_\text{CG}$ which
still implies spin-chirality decoupling.

In view of this controversy over the nature of the transition in the
three-dimensional Heisenberg spin glass, which is of great importance
for the understanding of spin glasses, 
we have undertaken a massive set of simulations to
study even larger sizes,\cite{sizes} $N\!=\!L^3$ where $L\!\le\!48$. Our
conclusion is that the data \textit{is} consistent with a common
transition temperature for spins and chiralities, though,
of course, numerics can never prove that they are \textit{exactly} equal.

The paper is organized as follows. In Sect.~\ref{sec:Model} we define the
model and the observables. Finite size scaling, which is central in our
analysis, is recalled in Sect.~\ref{sec:FSS}. Simulation details are in
Sect.~\ref{sec:Simulation}, while our equilibration tests are addressed in
Sect.~\ref{sec:Equilibration}. We find that a uniform allocation of
computational resources is inefficient (equilibration is much harder to
achieve for some particular samples).  The numerical results are in
Sect.~\ref{sec:Results}, while our conclusions are presented in
Sect.~\ref{sec:Conclusions}.

\section{Model and Observables}\label{sec:Model}
We use the standard Edwards-Anderson spin glass model on a cubic lattice 
\begin{equation}
{\cal H} = -\sum_{\langle i, j \rangle} J_{ij}\, {\bf S}_i \cdot
{\bf S}_j,
\end{equation}
where the ${\bf S}_i$ are 3-component classical
vectors of unit length at the sites of a simple cubic lattice, and the
$J_{ij}$ are nearest neighbor interactions with a Gaussian distribution
with zero mean and standard deviation unity.  Periodic boundary
conditions are applied.

The spin glass order parameter is $q^{\mu\nu}_i = S_i^{\mu(1)}
S_i^{\nu(2)}$, where ``(1)" and ``(2)'' are 
two identical 
copies of the system (same interactions), and
$\nu$ and $\mu$
are spin components. Its Fourier transform at wave vector ${\bfk}$,
is denoted by $\hat q^{\mu\nu}({\bfk})$.

For the Heisenberg spin-glass, Kawamura~\cite{kawamura:98} defines the
{\em chirality} from spins on a line: $ \kappa_i^\mu = {\bf
S}_{i+\hat{\mu}} \cdot ({\bf S}_i \times {\bf S}_{i-\hat{\mu}})$, where here
$\mu$ refers to a direction on the lattice.
The related
chiral spin-glass order parameter is $q_{\text{CG},i}^\mu =
\kappa_i^{\mu(1)}\kappa_i^{\mu(2)}$, its Fourier transform being $\hat
q^\mu_\text{CG}({\bfk})$.

The wave vector dependent susceptibilities are computed from the two
order parameters:
\begin{eqnarray}
\chi_\text{SG}({\bfk}) & = & N \sum_{\mu,\nu} [\langle
\left|\hat q^{\mu\nu}({\bfk})\right|^2 \rangle ]\av\, ,\\
\chi_\text{CG}^{\mu}({\bfk}) & = &
N [\langle\left| \hat q_\text{CG}^\mu({\bfk})\right|^2
\rangle ]\av\,,
\end{eqnarray}
where $\langle \cdots \rangle$ denotes a thermal
average and $[\cdots ]\av$ denotes an average over disorder.

The susceptibilities yield the second-moment finite-lattice estimator
of the correlation
length:\cite{cooper:82,palassini:99b,ballesteros:00,amit:05}
\begin{equation}
\xi_L = \frac{1}{2 \sin(\pi/L)} \,\,
\sqrt{\frac{\chi(0)}{\chi({\bfk}_\text{min})}-1} \,
,
\end{equation}
with 
${\bfk}_\text{min}\! =\!  (2\pi/L, 0, 0)$.
The spin and chiral~\cite{details-xicl}
correlation lengths are denoted by
$\xi_{\text{SG},L}$ and
$\xi_{\text{CG},L}$ respectively.

We also consider the spin and chiral  Binder ratios defined by
\begin{equation}
g_\text{SG}=\frac{11}{2}-\frac{9\,[\langle q^4 \rangle]\av}{2\,[\langle q^2\rangle]\av^2}\;,\quad
g_\text{CG}=\frac{5}{2} -\frac{3\,[\langle q_\text{CG}^4 \rangle]\av}
{2\,[\langle q_\text{CG}^2\rangle]\av^2}\;,
\label{g}
\end{equation}
\noindent where
$q^2 = \sum_{\mu\nu} \left[\hat q^{\mu\nu} ({\bfk}\!=\! 0) \right]^2$,
$q_\text{CG}^2 =
\sum_\mu \left[\hat q_\text{CG}^\mu ({\bfk}\!=\! 0) \right]^2$.

\section{Finite-Size Scaling}\label{sec:FSS}

Finite size scaling is a most useful data analysis method, that
exploits finite size effects where they are largest (at criticality)
to gather information on the {\em infinite} system, see
e.g.~Ref.~\onlinecite{amit:05}.

Finite size scaling takes the form of an asymptotic expansion on the
system size, $L$. To leading order, for a quantity $O$ diverging in the
thermodynamic limit as $O\propto |T-T_\text{c}|^{x_O}$, it takes the form
\begin{equation}
O(L,T)=L^{x_0/\nu} f\left(L^{1/\nu}(T-T_\mathrm{c})\right)\,,\label{fss-ansatz}
\end{equation}
where $f$ is an analytic function of its argument. In particular,
since $\xi_{\text{SG},L}/L$ is dimensionless it has the finite size
scaling form~\cite{binder:81b,ballesteros:96a,ballesteros:00,lee:03}
\begin{equation}
\frac{\xi_{\text{SG},L}}{L} = \widetilde{X}\left(L^{1/\nu}(T - T_\text{SG})\right) ,
\label{fss}
\end{equation} 
where $\nu$ is the correlation length exponent.
There are similar expressions for
$\xi_{\text{CG}, L}/L$, 
and also for the Binder ratios in Eq.~(\ref{g}) since these
too are dimensionless.

We shall see
that corrections to scaling are quite
large, even for the large sizes that we study, and so we need to
consider \textit{corrections}
to the asymptotic scaling form in Eq.~(\ref{fss}). To investigate this
we determine the intersection temperatures
$T_\text{SG}^\star(L, s L)$, where the data for $\xi_{\text{SG},L}/L$
agree for sizes $L$ and $s L$, i.e.
\begin{equation}
\label{quotient}
\frac{\xi_{\text{SG},L}}{ L} = \frac{\xi_{\text{SG},s L}}{ s L} \, ,
\end{equation}
with an analogous expression for the chiral data.

Whereas Eq.~(\ref{fss}) predicts that all the $T_\text{SG}^\star(L, s
L)$ are equal to the spin glass transition temperature $T_\text{SG}$,
when one
includes the leading corrections to scaling
the $T_\text{SG}^\star(L, s L)$ are given 
by~\cite{binder:81b,ballesteros:96a,hasenbusch:08b}
\begin{equation}
T_\text{SG}^\star(L, s L) - T_\text{SG} = 
A_\text{SG}^{(s)}\, L^{-\omega-\frac{1}{\nu}}\, .
\label{Tstar}
\end{equation}
Here, $\omega$ is the
exponent for the leading correction to scaling while the amplitude is
\begin{equation}
A_\text{SG}^{(s)}= A_\text{SG}\,\frac{1 - s^{-\omega}}{s^{1/\nu} - 1}\, ,
\end{equation}
with $A_\text{SG}$ a (non-universal) constant. In practice, we do not
have enough information to determine the $s$ dependence in
Eq.~(\ref{Tstar}), so we take the $A_\text{SG}^{(s)}$
to be separate constants for each value of $s$ that we use
($s = 2$ and $3/2$).

In fact we may combine Eqs. \eqref{fss-ansatz} and \eqref{quotient},
to obtain a modern form of Nightingale's phenomenological renormalization,\cite{nightingale:76} the so called quotient method:\cite{ballesteros:96a}
\begin{equation}
\frac{O\left (sL,T_\text{SG}^\star(L, s L)\right)}
{O\left (L,T_\text{SG}^\star(L, s L)\right)}=s^{x_O/\nu_\text{SG}}[1 + \tilde A_{O,s} L^{-\omega}+\ldots]\,,\label{QUOTIENTS}
\end{equation}
where the dots stand for higher order scaling corrections. Were
$T_\text{CG}$ and $T_\text{SG}$ to be different, a similar expression
would hold for quantities diverging at $T_\text{CG}$. In particular, one
may use Eq.~\eqref{QUOTIENTS} with temperature derivatives (to obtain
$1+1/\nu$) or with the susceptibilities at zero wave number (to obtain
$2-\eta$, where $\eta$ is the anomalous dimension).

\section{Simulation Details}\label{sec:Simulation}

Simulations are run in parallel on $N_T$ processors at $N_T$ different
temperatures using the parallel tempering~\cite{hukushima:96} (PT)
method to speed up equilibration at low-$T$, see
Table~\ref{tab:params} for the parameters.

As discussed in other
work,\cite{campos:06,lee:07,viet:09} three types of moves are
performed: (i) ``overrelaxation'' (OR) sweeps (which do not change the
energy), (ii) ``heat-bath'' (HB) sweeps (which do change the energy),
and (iii) parallel tempering (PT) sweeps in which the spin
configurations at neighboring temperatures are swapped with a
probability which satisfies the detailed balance condition. It is
important to include OR sweeps because not only is
the code for them much simpler (and hence faster) than that for the
HB sweeps, but also because OR moves are very
efficient~\cite{campos:06,lee:07,pixley:08} in equilibrating the system,
so many \textit{fewer} sweeps are required than in a simulation with
only HB and PT moves.  Nonetheless, a fraction of  the moves must be HB in
order to change the energy.
Because of the PT moves, 
the temperature of a set of spins (a ``copy") is not fixed but does a
random walk between the minimum and maximum temperature in the set.
In this way, each copy of spins can visit 
different ``valleys" in configuration space
with the correct statistical weight, even at low
temperature.

\begin{table}[!tb]
\caption{
Parameters of the simulations. $N_T$ is the  number of temperatures. For
the larger sizes, the number of sweeps varied from sample to sample. We
show values for
$N_\text{sweep}^\text{min}$ and $N_\text{sweep}^\text{max}$, the
minimum and maximum number of overrelaxation sweeps.
}
\label{tab:params}
\begin{tabular*}{\columnwidth}{@{\extracolsep{\fill}}rllrc}
\hline
\hline
$L$ & $N_\text{sweep}^\text{min}$  & $N_\text{sweep}^\text{max}$  &
$N_T$ & $N_\text{samp}$ \\
\hline
8    & $5.0 \times 10^5$ & $5.0 \times 10^5$ & 5  & 984  \\
12    & $7.5 \times 10^5$ & $7.5 \times 10^5$ & 9  & 984  \\
16    & $1.0 \times 10^6$ & $1.0 \times 10^6$ & 15 & 984  \\
24    & $1.5 \times 10^6$ & $1.2 \times 10^7$ & 27 & 984  \\
32    & $4.0 \times 10^6$ & $1.2 \times 10^8$ & 43 & 984  \\
48    & $6.0 \times 10^7$ & $6.0 \times 10^8$ & 79 & 164   \\
\hline
\hline
\end{tabular*}
\end{table}

The $N_T$ temperatures were arranged in a
geometric progression
between $T_\text{min} = 0.12$ and $T_\text{max} = 0.19$.
We do
1 HB sweep followed by $5L/4$ OR sweeps and then
$100$ PT sweeps. We found a net CPU gain by doing a number of OR sweeps
between HB sweeps which is somewhat greater than $L$,
perhaps because this transfers a fluctuation right 
across the system. We do a
large number (100) of PT sweeps following right after each other,
because the PT sweeps are very inexpensive in CPU
time.

\begin{figure}
\includegraphics[height=\columnwidth,angle=270,trim=12 35 16 30]{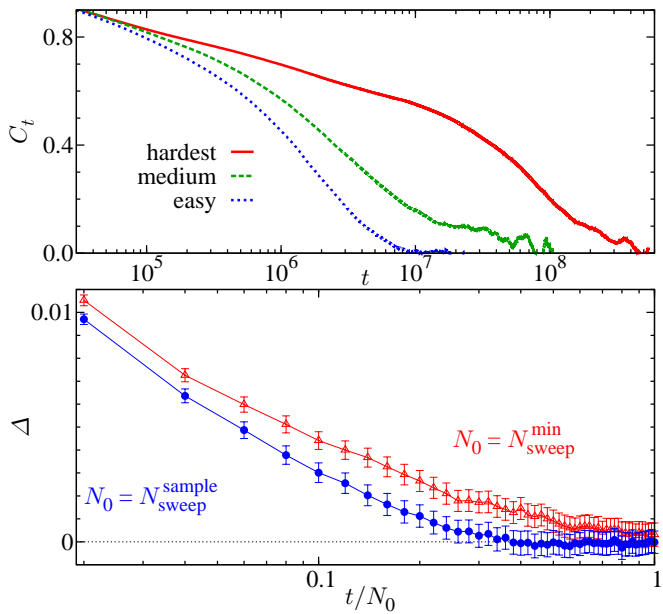}
\caption{(color online) {\bf Top:} Parallel tempering autocorrelation
  function,\cite{details-tau} as computed for three representative
  $L\!=\!48$ samples. Here {\em easy} means that after
  $N_\text{sweep}^\text{min}$ MC steps, see Table~\ref{tab:params},
  the equilibration criterion was met (42\% of samples), while {\em
    medium} samples (34\%) required up to $2\,
  N_\text{sweep}^\text{min}$ MC steps.  {\bf Bottom:} The quantity
  $\varDelta$ in Eq.~(\ref{equiltest}) as a function of MC time, both in
  units of $N_\text{sweep}^\text{min}$ (red triangles) and in units of
  the maximum number of sweeps for each sample (blue circles). For the latter,
  note  that the data is computed at different times for different samples.}
\label{fig:equil}
\end{figure}

\section{Equilibration}\label{sec:Equilibration}

We do several tests to ensure
equilibration. Firstly, we require that data satisfy the
relation~\cite{katzgraber:01,lee:07}
\begin{equation}
\varDelta \equiv \left[
\frac{q_\text{s} - q_\text{l}}{T} 
+  \frac{2}{z}\, U \right]\av= 0, 
\label{equiltest}
\end{equation}
which is valid for a Gaussian bond distribution. Here
\begin{eqnarray}
U & = & - \sum_{\langle i, j \rangle} J_{ij} \langle {\bf S}_i \cdot 
{\bf S}_j \rangle \, , \\
q_\text{l} & = & (1/N_\text{b})\sum_{\langle i, j \rangle}
\langle
{\bf S}_i \cdot {\bf S}_j \rangle^2 \, ,\\
q_\text{s} & = &
(1/N_\text{b})\sum_{\langle i, j \rangle}\langle ({\bf S}_i \cdot {\bf
S}_j)^2 \rangle  \, ,
\end{eqnarray}
in which $U$ is the thermally averaged energy per spin, $q_\text{l}$
is called the ``link overlap'', $N_\text{b} = (z/2)N$ is the number of
nearest neighbor bonds, and $z\ (=6\ \mbox{here})$ is the lattice
coordination number.  Both $U$ and $q_\text{s}$, being a single
thermal average, come close to equilibrium relatively quickly as the
number of MC sweeps increases. However, $q_\text{l}$ involves a double
thermal average, which is determined from two separate copies
initialized with random spin configurations, and hence is initially
very small.  As the simulation proceeds, $q_\text{l}$ increases
towards its equilibrium value, so $\varDelta$ in Eq.~(\ref{equiltest})
will initially be positive but will become zero (and stay zero) when
equilibrium is reached. Data is shown in Fig.~\ref{fig:equil} for
$L=48$ (the largest size) at $T=0.12$ (the lowest temperature).

Eq.~(\ref{equiltest}) also provides a {\em control
  variate}~\cite{fernandez:09} to reduce statistical errors in
$\xi_{\text{SG},L}$ and $\xi_{\text{CG},L}$. The key is in the strong
statistical correlations between the Monte Carlo estimator for
$\varDelta$ and those for the susceptibilities. Since we happen to
know that $\varDelta=0$, reduced-variance estimators for the
susceptibilities are obtained straightforwardly (see
Ref.~\onlinecite{fernandez:09} for details). In practice, this method
halves the errors for $\xi_\text{CG,SG}$ at $T=0.12$ (however, for $T
\gtrsim 0.14$ the gain is less than $10\%$).

Some samples are harder
to equilibrate than others so, ideally, we should spend more MC sweeps on
the ``hard'' ones than on ``easy'' ones. The key to classifying samples in
a PT simulation is to consider the dynamics of the temperature
random-walk.
In ``hard'' samples the $T$
random-walk is slower (a copy trapped in a deep valley needs a longer
time to wander to a $T$ high enough to escape). We use correlation
functions and autocorrelation times to formalize this idea, see
Fig.~\ref{fig:equil} and the comments in Ref.~\onlinecite{details-tau}.

For each sample, we impose a minimum number of sweeps,
Table~\ref{tab:params}, then keep simulating until the total number of
MC iterations exceed 9 autocorrelation times. For $L\!=\!48$, the
average number of MC iterations per sample was 1.8 times the minimum.
Figure~\ref{fig:equil} shows that the data equilibrates more
convincingly by running the ``hard'' samples for longer than the
``easy'' samples.

\begin{figure}
\includegraphics[height=\columnwidth,angle=270,trim=0 10 0 30]{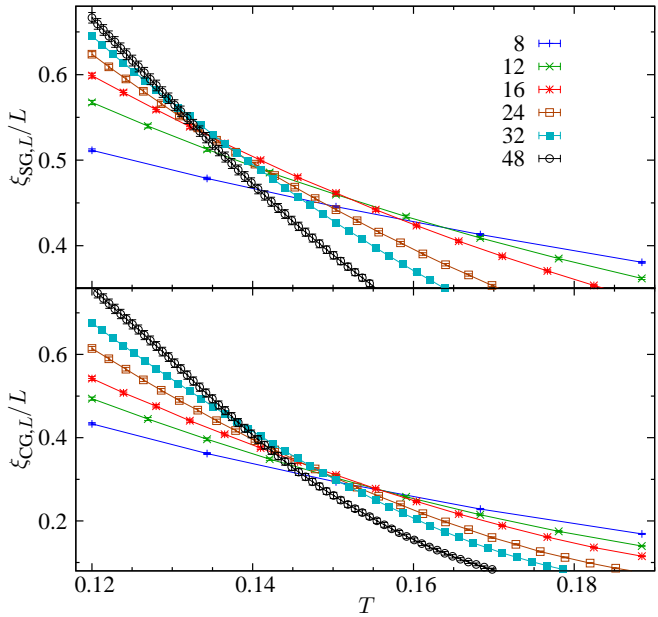}
\caption{(color online)
Data for the spin glass and chiral glass 
correlation lengths divided by system size. For $L
\to\infty$ the data should intersect at 
the transition temperature. Here, the data does not show a common
intersection temperature, indicating that there are 
strong corrections to scaling for the range of sizes studied.
}
\label{fig:xi_L}
\end{figure}

\section{Results}\label{sec:Results}
We now present our results. Figure \ref{fig:xi_L} shows data for the
spin glass and chiral glass correlation lengths divided by $L$. The
resulting intersection temperatures, obtained from a jackknife
analysis, are shown in Table~\ref{tab:Tstars}.

\begin{table}[!tb]
\caption{ Table of intersection temperatures $T^\star(L, s L)$ for the
spin glass and chiral glass correlation length data presented in
Fig.~\ref{fig:xi_L}.  Also shown are estimates for the exponents
$\nu$ and $2-\eta$ for the case of $s=2$, using the quotient
method,\cite{ballesteros:96a,amit:05} see Eq.~(\ref{QUOTIENTS}). The
operators used are $\partial_T \xi_{\text{SG},L}\,$,
$\partial_T \xi_{\text{CG},L}\,$ $\chi_\text{SG}$ and
$\chi_\text{CG}$ which have scaling exponents $1+1/
{\nu_\text{SG}}$, \hbox{$1+1 / \nu_\text{CG}$}, $2-\eta_\text{SG}$ and
$2-\eta_\mathrm{CG}$ respectively.  Spin (chiral) exponents are computed
from data at $T^\star_\text{SG}(L, s L)$ ($T^\star_\text{CG}(L, s L)$).  }
\label{tab:Tstars}

\begin{tabular*}{\columnwidth}{@{\extracolsep{\fill}}rrccllcc}
\hline
\hline
{$L$} & $s L$ & $T^\star_\text{SG}$ & $ T^\star_\text{CG}$
&\multicolumn1{c}{$\nu_\text{SG}$} & \multicolumn1{c}{$\nu_\text{CG}$} & $2-\eta_\text{SG}$ & $2-\eta_\text{CG}$ \\
\hline
8   &  16 & 0.158(1) & 0.156(1) &  1.01(2) & 1.34(5)&  1.99(1) & 0.72(2)\\
12  &  24 & 0.142(2) & 0.150(1) &  1.35(5) & 1.51(6)&  2.08(1) & 0.96(3)\\
16  &  32 & 0.136(1) & 0.147(1) &  1.50(7) & 1.46(6)&  2.14(1) & 1.11(3)\\
24  &  48 & 0.133(2) & 0.142(1) &  1.49(13)& 1.30(8)&  2.19(2) & 1.44(4)\\ 
8   &  12 & 0.164(2) & 0.157(2) &          &        &          &        \\
16  &  24 & 0.135(2) & 0.147(2) &          &        &          &        \\
32  &  48 & 0.130(3) & 0.138(2) &          &        &          &        \\
\hline
\hline
\end{tabular*} 
\end{table}

\begin{figure}
\includegraphics[height=\columnwidth,angle=270,trim=15 25 0 40]{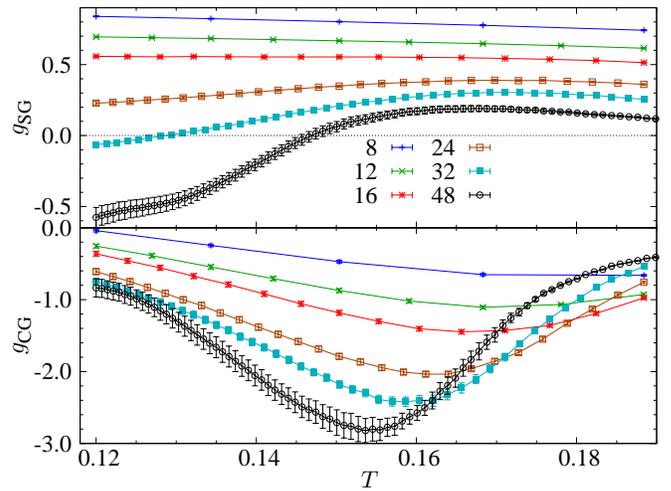}
\caption{(color on line)
Data for the spin glass and chiral glass Binder ratios defined in
Eq.~(\ref{g}) of the text.
}
\label{fig:g}
\end{figure}

Since Eq.~(\ref{Tstar}) holds only asymptotically, for large
$L$, it is necessary to decide on the smallest size
$L_\text{min}$ to be included in the analysis.
We consider first the five pairs of sizes with
$L_\text{min}=12$, see Table~\ref{tab:Tstars}.  Fitting spin and
chiral data separately there are 4 parameters for each:
$T_{\text{CG},\text{SG}},$ the exponent $\omega + 1/\nu$, and amplitudes
$A_{\text{CG},\text{SG}}^{(2)}$ and $A_{\text{CG},\text{SG}}^{(3/2)}$ for
the $s=2$ and $s=3/2$ size ratios.  We determine the best fit parameters, and
estimate the quality of the fit from\cite{details-chi2} the value
of $\chi^2$.
Fitting the spin data to
Eq.~(\ref{Tstar}), gives $T_\text{SG} = 0.129^{+0.003}_{-0.016}$,
which is compatible
with Viet and Kawamura's result of $0.120(6)$.  However, in the
chiral sector $\chi^2$ as a
function of $T_\text{CG}$ does not have a local minimum with
$T_{\text{CG}}>0$, so subleading scaling corrections are sizable for
chiralities and $L_\text{min}=12$.

Hence 
we have also performed an analysis with a larger value,
$L_\text{min}\!=\!16$. Unfortunately, we only have data for four pairs of sizes,
and still four parameters to be fitted if we fit the spin and chiral
data separately. Since the number of points is equal to the number of
parameters we do not gain useful information. However, if we assume a
\textit{common transition temperature} and do a joint fit we have 8
data points, and 6 parameters (1 transition temperature, one exponent,
and 4 amplitudes). The resulting fit gives $T_\text{c}\, (=\!T_\text{SG}
\!=\!T_\text{CG}) =
0.120^{+0.010}_{-0.100}$,
with a $\chi^2$ per degree of freedom of 0.029 so the fit is good. 
The error bar on $T_\text{c}$ is very large on the low-$T$ side but if we
assume that the exponent $1/\nu + \omega$ in Eq.~(\ref{Tstar}) is greater
than 0.5, plausible given the values for $\nu$ in Table \ref{tab:Tstars}, 
we find $T_\text{c} = 0.120^{+0.010}_{-0.004}$,
much more tightly constrained.

In Fig.~\ref{fig:g} we show data for the spin glass and chiral glass
Binder ratios defined in Eq.~(\ref{g}). Our definition of $g_\text{CG}$
differs from that of Kawamura and Viet,\cite{viet:09} and our results
have an even more pronounced negative dip. Interestingly, we find that
the results for $g_\text{SG}$ \textit{also} become negative at the
largest sizes. Hence, the apparent vanishing of
$g_\text{SG}$ near $T_\text{CG}$, a strong argument for spin-chirality
decoupling,\cite{hukushima:05} is an artifact caused by the
lattice sizes being too small and the temperatures too high.
Our interpretation
of the Binder parameter data is that there is negative dip in
both channels, and
the minimum of this dip approaches the 
transition temperature as $L$
grows. The chiral-dip approaches $T_\text{CG}$ from high temperatures,
while the
spin-dip approaches $T_\text{SG}$ from low temperatures, where plausibly 
$T_\text{SG} = T_\text{CG}$. However, much larger sizes would be needed
to confirm this hypothesis.

By studying sizes $L \le 32$ Viet and Kawamura~\cite{viet:09,viet:09a}
find $T_\text{SG} = 0.120(6)$ and $T_\text{CG} = 0.145(4)$. Since the
difference is about 3.5 times the errors they argue that $T_\text{CG} >
T_\text{SG}$. However, their value for $T_\text{CG}$ is actually higher
than our intersection temperatures for $L = 48$ shown in
Table~\ref{tab:Tstars}, and so seems to us to be too high. Also they
estimate the transition temperatures from $T^\star(L, s L) -
T_\text{CG} = {\rm const.} / L_\text{av}$, where $L_\text{av}$ is the
average of $L$ and $s L$, rather than Eq.~(\ref{Tstar}). In other words
they replace the exponent $\omega + 1/\nu$ by 1, and the $s$
dependence in Eq.~(\ref{Tstar}) by $2/(1+s)$. At the very least, 
we argue that these replacements lead to an underestimate of
the error bars.
Hence, we do not feel that the results of Viet and Kawamura contradict
our conclusion that the data is consistent with the spin and chiral
glass transition temperatures being equal.

\section{Conclusions}\label{sec:Conclusions}
In summary, our low-temperature simulations for the Heisenberg spin
glass are unprecedented in system size.  To achieve thermalization, we
have needed not only a huge amount of CPU ($7\times 10^6$ hours) but a
careful sample-by-sample thermalization check that allowed us to
concentrate efforts on the ``hard'' samples. The results for the
spin-glass sector can be accounted for using only leading-order scaling
corrections, but subleading corrections are sizable for the chiral
glass sector. This is the reason for the overestimate of $T_\text{CG}$
in previous work.\cite{viet:09,viet:09a}  Data for $L\geq 16$ support
the most economic scenario, $T_\text{SG}\!=\!T_\text{CG}$.  We also
see that
the spin Binder parameter is {\em
not} trivial at $T_\text{CG}$.

\begin{acknowledgments}
The CPU time used was $1.3 \times 10^6$ hours at 
CINECA (through the EU DEISA initiative), $4.7 \times 10^6$ hours at
the Mare Nostrum, $10^6$ hours at Caesaraugusta and $10^5$ hours at the 
Hierarchical Systems
Research Foundation.  We were partly supported by MICINN (Spain,
research contract No. FIS2006-08533-C03).  The authors thankfully
acknowledge the computer resources, technical expertise and assistance
provided by the staff at the {\em Red Espa\~nola de
Supercomputaci\'on--Barcelona Supercomputing Center\/}, and at 
CINECA.
\end{acknowledgments}

\bibliography{refs,comments}

\begin{thebibliography}{26}
\expandafter\ifx\csname natexlab\endcsname\relax\def\natexlab#1{#1}\fi
\expandafter\ifx\csname bibnamefont\endcsname\relax
  \def\bibnamefont#1{#1}\fi
\expandafter\ifx\csname bibfnamefont\endcsname\relax
  \def\bibfnamefont#1{#1}\fi
\expandafter\ifx\csname citenamefont\endcsname\relax
  \def\citenamefont#1{#1}\fi
\expandafter\ifx\csname url\endcsname\relax
  \def\url#1{\texttt{#1}}\fi
\expandafter\ifx\csname urlprefix\endcsname\relax\def\urlprefix{URL }\fi
\providecommand{\bibinfo}[2]{#2}
\providecommand{\eprint}[2][]{\url{#2}}

\bibitem[{\citenamefont{Ballesteros et~al.}(2000)\citenamefont{Ballesteros,
  Cruz, Fernandez, Martin-Mayor, Pech, Ruiz-Lorenzo, Tarancon, Tellez, Ullod,
  and Ungil}}]{ballesteros:00}
\bibinfo{author}{\bibfnamefont{H.~G.} \bibnamefont{Ballesteros}},
  \bibinfo{author}{\bibfnamefont{A.}~\bibnamefont{Cruz}},
  \bibinfo{author}{\bibfnamefont{L.~A.} \bibnamefont{Fernandez}},
  \bibinfo{author}{\bibfnamefont{V.}~\bibnamefont{Martin-Mayor}},
  \bibinfo{author}{\bibfnamefont{J.}~\bibnamefont{Pech}},
  \bibinfo{author}{\bibfnamefont{J.~J.} \bibnamefont{Ruiz-Lorenzo}},
  \bibinfo{author}{\bibfnamefont{A.}~\bibnamefont{Tarancon}},
  \bibinfo{author}{\bibfnamefont{P.}~\bibnamefont{Tellez}},
  \bibinfo{author}{\bibfnamefont{C.~L.} \bibnamefont{Ullod}}, \bibnamefont{and}
  \bibinfo{author}{\bibfnamefont{C.}~\bibnamefont{Ungil}},
  \bibinfo{journal}{Phys. Rev. B} \textbf{\bibinfo{volume}{62}},
  \bibinfo{pages}{14237} (\bibinfo{year}{2000}),
  \eprint{(arXiv:cond-mat/0006211)}.

\bibitem[{\citenamefont{Katzgraber et~al.}(2006)\citenamefont{Katzgraber,
  K\"orner, and Young}}]{katzgraber:06}
\bibinfo{author}{\bibfnamefont{H.~G.} \bibnamefont{Katzgraber}},
  \bibinfo{author}{\bibfnamefont{M.}~\bibnamefont{K\"orner}}, \bibnamefont{and}
  \bibinfo{author}{\bibfnamefont{A.~P.} \bibnamefont{Young}},
  \bibinfo{journal}{Phys. Rev. B} \textbf{\bibinfo{volume}{73}},
  \bibinfo{pages}{224432} (\bibinfo{year}{2006}),
  \eprint{(arXiv:cond-mat/0602212)}.

\bibitem[{\citenamefont{Hasenbusch et~al.}(2008)\citenamefont{Hasenbusch,
  Pelissetto, and Vicari}}]{hasenbusch:08b}
\bibinfo{author}{\bibfnamefont{M.}~\bibnamefont{Hasenbusch}},
  \bibinfo{author}{\bibfnamefont{A.}~\bibnamefont{Pelissetto}},
  \bibnamefont{and} \bibinfo{author}{\bibfnamefont{E.}~\bibnamefont{Vicari}},
  \bibinfo{journal}{Phys. Rev. B} \textbf{\bibinfo{volume}{78}},
  \bibinfo{pages}{214205} (\bibinfo{year}{2008}), \eprint{(arXiv:0809.3329)}.

\bibitem[{\citenamefont{Omari et~al.}(1983)\citenamefont{Omari, Prejean, and
  Souletie}}]{omari:83}
\bibinfo{author}{\bibfnamefont{R.}~\bibnamefont{Omari}},
  \bibinfo{author}{\bibfnamefont{J.~J.} \bibnamefont{Prejean}},
  \bibnamefont{and} \bibinfo{author}{\bibfnamefont{J.}~\bibnamefont{Souletie}},
  \bibinfo{journal}{J. de Physique} \textbf{\bibinfo{volume}{44}},
  \bibinfo{pages}{1069} (\bibinfo{year}{1983}).

\bibitem[{\citenamefont{Kawamura}(1998)}]{kawamura:98}
\bibinfo{author}{\bibfnamefont{H.}~\bibnamefont{Kawamura}},
  \bibinfo{journal}{Phys. Rev. Lett.} \textbf{\bibinfo{volume}{80}},
  \bibinfo{pages}{5421} (\bibinfo{year}{1998}),
  \eprint{(arXiv:cond-mat/9805117)}.

\bibitem[{\citenamefont{Hukushima and Kawamura}(2005)}]{hukushima:05}
\bibinfo{author}{\bibfnamefont{K.}~\bibnamefont{Hukushima}} \bibnamefont{and}
  \bibinfo{author}{\bibfnamefont{H.}~\bibnamefont{Kawamura}},
  \bibinfo{journal}{Phys. Rev. B} \textbf{\bibinfo{volume}{72}},
  \bibinfo{pages}{144416} (\bibinfo{year}{2005}).

\bibitem[{\citenamefont{Lee and Young}(2003)}]{lee:03}
\bibinfo{author}{\bibfnamefont{L.~W.} \bibnamefont{Lee}} \bibnamefont{and}
  \bibinfo{author}{\bibfnamefont{A.~P.} \bibnamefont{Young}},
  \bibinfo{journal}{Phys. Rev. Lett.} \textbf{\bibinfo{volume}{90}},
  \bibinfo{pages}{227203} (\bibinfo{year}{2003}),
  \eprint{(arXiv:cond-mat/0302371)}.

\bibitem[{\citenamefont{Campos et~al.}(2006)\citenamefont{Campos, Cotallo-Aban,
  Martin-Mayor, Perez-Gaviro, and Tarancon}}]{campos:06}
\bibinfo{author}{\bibfnamefont{I.}~\bibnamefont{Campos}},
  \bibinfo{author}{\bibfnamefont{M.}~\bibnamefont{Cotallo-Aban}},
  \bibinfo{author}{\bibfnamefont{V.}~\bibnamefont{Martin-Mayor}},
  \bibinfo{author}{\bibfnamefont{S.}~\bibnamefont{Perez-Gaviro}},
  \bibnamefont{and} \bibinfo{author}{\bibfnamefont{A.}~\bibnamefont{Tarancon}},
  \bibinfo{journal}{Phys. Rev. Lett.} \textbf{\bibinfo{volume}{97}},
  \bibinfo{pages}{217204} (\bibinfo{year}{2006}),
  \eprint{(arXiv:cond-mat/0605327)}.

\bibitem[{\citenamefont{Lee and Young}(2007)}]{lee:07}
\bibinfo{author}{\bibfnamefont{L.~W.} \bibnamefont{Lee}} \bibnamefont{and}
  \bibinfo{author}{\bibfnamefont{A.~P.} \bibnamefont{Young}},
  \bibinfo{journal}{Phys. Rev. B} \textbf{\bibinfo{volume}{76}},
  \bibinfo{pages}{024405} (\bibinfo{year}{2007}),
  \eprint{(arXiv:cond-mat/0703770)}.

\bibitem[{\citenamefont{Viet and Kawamura}(2009{\natexlab{a}})}]{viet:09}
\bibinfo{author}{\bibfnamefont{D.~X.} \bibnamefont{Viet}} \bibnamefont{and}
  \bibinfo{author}{\bibfnamefont{H.}~\bibnamefont{Kawamura}},
  \bibinfo{journal}{Phys. Rev. Lett.} \textbf{\bibinfo{volume}{102}},
  \bibinfo{pages}{027202} (\bibinfo{year}{2009}{\natexlab{a}}),
  \eprint{(arXiv:0808.3328)}.

\bibitem[{\citenamefont{Viet and Kawamura}(2009{\natexlab{b}})}]{viet:09a}
\bibinfo{author}{\bibfnamefont{D.~X.} \bibnamefont{Viet}} \bibnamefont{and}
  \bibinfo{author}{\bibfnamefont{H.}~\bibnamefont{Kawamura}}
  (\bibinfo{year}{2009}{\natexlab{b}}),
  \bibinfo{note}{arXiv:cond-mat/0904.3699}.

\bibitem[{siz()}]{sizes}
\bibinfo{note}{To our knowledge, a $48^3$ lattice is the largest spin glass
  that has been thermalized near a finite temperature phase transition. It is
  curious that larger sizes can be studied for the Heisenberg model than for
  the Ising cases (for which the $28^3$ samples studied by Hasenbusch et
  al.~\cite{hasenbusch:08b} seems to be the record), even though the updating
  code is more complicated. Evidently, the barriers between ``valleys'' are
  lower in the Heisenberg model.}

\bibitem[{\citenamefont{Cooper et~al.}(1982)\citenamefont{Cooper, Freedman, and
  Preston}}]{cooper:82}
\bibinfo{author}{\bibfnamefont{B.}~\bibnamefont{Cooper}},
  \bibinfo{author}{\bibfnamefont{B.}~\bibnamefont{Freedman}}, \bibnamefont{and}
  \bibinfo{author}{\bibfnamefont{D.}~\bibnamefont{Preston}},
  \bibinfo{journal}{Nucl. Phys. B.} \textbf{\bibinfo{volume}{210}},
  \bibinfo{pages}{210} (\bibinfo{year}{1982}).

\bibitem[{\citenamefont{Palassini and Caracciolo}(1999)}]{palassini:99b}
\bibinfo{author}{\bibfnamefont{M.}~\bibnamefont{Palassini}} \bibnamefont{and}
  \bibinfo{author}{\bibfnamefont{S.}~\bibnamefont{Caracciolo}},
  \bibinfo{journal}{Phys. Rev. Lett.} \textbf{\bibinfo{volume}{82}},
  \bibinfo{pages}{5128} (\bibinfo{year}{1999}),
  \eprint{(arXiv:cond-mat/9904246)}.

\bibitem[{\citenamefont{Amit and Matin-Mayor}(2005)}]{amit:05}
\bibinfo{author}{\bibfnamefont{D.}~\bibnamefont{Amit}} \bibnamefont{and}
  \bibinfo{author}{\bibfnamefont{V.}~\bibnamefont{Matin-Mayor}},
  \emph{\bibinfo{title}{Field Theory, the Renormalization Group and Critical
  Phenomena}} (\bibinfo{publisher}{World Scientific},
  \bibinfo{address}{Singapore}, \bibinfo{year}{2005}).

\bibitem[{det({\natexlab{a}})}]{details-xicl}
\bibinfo{note}{For the CG, one considers a transverse or parallel $\xi_{CG,L}$,
  depending on whether $\hat\mu\cdot\bfk_\text{min}=0$ or
  not~\cite{lee:03,campos:06}. We report only the parallel $\xi_{\text{CG},L}$
  as the two coincide within errors.}

\bibitem[{\citenamefont{Binder}(1981)}]{binder:81b}
\bibinfo{author}{\bibfnamefont{K.}~\bibnamefont{Binder}}, \bibinfo{journal}{Z.
  Phys. B} \textbf{\bibinfo{volume}{43}}, \bibinfo{pages}{119}
  (\bibinfo{year}{1981}).

\bibitem[{\citenamefont{Ballesteros et~al.}(1996)\citenamefont{Ballesteros,
  Fernandez, Martin-Mayor, Pech, and Mu\~noz Sudupe}}]{ballesteros:96a}
\bibinfo{author}{\bibfnamefont{H.~G.} \bibnamefont{Ballesteros}},
  \bibinfo{author}{\bibfnamefont{L.~A.} \bibnamefont{Fernandez}},
  \bibinfo{author}{\bibfnamefont{V.}~\bibnamefont{Martin-Mayor}},
  \bibinfo{author}{\bibfnamefont{J.}~\bibnamefont{Pech}}, \bibnamefont{and}
  \bibinfo{author}{\bibfnamefont{A.}~\bibnamefont{Mu\~noz Sudupe}},
  \bibinfo{journal}{Phys. Lett. B} \textbf{\bibinfo{volume}{387}},
  \bibinfo{pages}{125} (\bibinfo{year}{1996}),
  \eprint{(arXiv:cond-mat/9606203)}.

\bibitem[{\citenamefont{Nightingale}(1976)}]{nightingale:76}
\bibinfo{author}{\bibfnamefont{M.~P.} \bibnamefont{Nightingale}},
  \bibinfo{journal}{Physica A} \textbf{\bibinfo{volume}{83}},
  \bibinfo{pages}{561} (\bibinfo{year}{1976}).

\bibitem[{\citenamefont{Hukushima and Nemoto}(1996)}]{hukushima:96}
\bibinfo{author}{\bibfnamefont{K.}~\bibnamefont{Hukushima}} \bibnamefont{and}
  \bibinfo{author}{\bibfnamefont{K.}~\bibnamefont{Nemoto}},
  \bibinfo{journal}{J. Phys. Soc. Japan} \textbf{\bibinfo{volume}{65}},
  \bibinfo{pages}{1604} (\bibinfo{year}{1996}),
  \eprint{(arXiv:cond-mat/9512035)}.

\bibitem[{\citenamefont{Pixley and Young}(2008)}]{pixley:08}
\bibinfo{author}{\bibfnamefont{J.}~\bibnamefont{Pixley}} \bibnamefont{and}
  \bibinfo{author}{\bibfnamefont{A.}~\bibnamefont{Young}},
  \bibinfo{journal}{Phys. Rev. B} \textbf{\bibinfo{volume}{78}},
  \bibinfo{pages}{014419} (\bibinfo{year}{2008}).

\bibitem[{det({\natexlab{b}})}]{details-tau}
\bibinfo{note}{Given the set of temperatures $\{T_i\}$, let $f(T)$ be a cubic
  polynomial in $T^{-1}$ (unique up to an irrelevant multiplicative constant)
  with $\sum_i f(T_i)\!=\!0$, $f'(T_\text{max})=0$ and changing sign at
  $0.14\approx T_\text{c}$. Let $f_t$ be $f(T)$ for the $T$ occupied by one
  copy of the system at time $t$. We first coarse grain $f_t$ by averaging over
  100 consecutive MC sweeps, then compute its autocorrelation function and
  integrated autocorrelation time (see e.g. Refs.
  \onlinecite{sokal:97,amit:05})}.

\bibitem[{\citenamefont{Katzgraber et~al.}(2001)\citenamefont{Katzgraber,
  Palassini, and Young}}]{katzgraber:01}
\bibinfo{author}{\bibfnamefont{H.~G.} \bibnamefont{Katzgraber}},
  \bibinfo{author}{\bibfnamefont{M.}~\bibnamefont{Palassini}},
  \bibnamefont{and} \bibinfo{author}{\bibfnamefont{A.~P.} \bibnamefont{Young}},
  \bibinfo{journal}{Phys. Rev. B} \textbf{\bibinfo{volume}{63}},
  \bibinfo{pages}{184422} (\bibinfo{year}{2001}),
  \eprint{(arXiv:cond-mat/0007113)}.

\bibitem[{\citenamefont{Fernandez and Martin-Mayor}(2009)}]{fernandez:09}
\bibinfo{author}{\bibfnamefont{L.}~\bibnamefont{Fernandez}} \bibnamefont{and}
  \bibinfo{author}{\bibfnamefont{V.}~\bibnamefont{Martin-Mayor}},
  \bibinfo{journal}{Phys. Rev. E} \textbf{\bibinfo{volume}{79}},
  \bibinfo{pages}{051109} (\bibinfo{year}{2009}), \eprint{(arXiv:0904.3415)}.

\bibitem[{det({\natexlab{c}})}]{details-chi2}
\bibinfo{note}{The $T^*_{\text{SG},\text{CG}}(L,sL)$ are statistically
  correlated so one should use the full covariance matrix to compute $\chi^2$.
  However, we find that considering only diagonal covariances does not
  significantly change the results. We give results obtained with diagonal
  covariances since these can be reproduced from the data in
  Table~\ref{tab:Tstars}.}

\bibitem[{\citenamefont{Sokal}(1997)}]{sokal:97}
\bibinfo{author}{\bibfnamefont{A.}~\bibnamefont{Sokal}}, in
  \emph{\bibinfo{booktitle}{Functional Integration: Basics and Applications}},
  edited by \bibinfo{editor}{\bibfnamefont{C.}~\bibnamefont{DeWitt-Morette}},
  \bibinfo{editor}{\bibfnamefont{P.}~\bibnamefont{Cartier}}, \bibnamefont{and}
  \bibinfo{editor}{\bibfnamefont{A.}~\bibnamefont{Folacci}}
  (\bibinfo{publisher}{Plenum}, \bibinfo{year}{1997}).

\end{thebibliography}

\end{document}